\documentclass[smallextended]{svjour3}     % onecolumn (second format)
\smartqed  % flush right qed marks, e.g. at end of proof
\usepackage{graphicx}
\usepackage{graphicx}
\usepackage{color}
\usepackage{rotating}
\begin{document}

\title{B-Pol: Detecting Primordial Gravitational Waves \\
Generated During Inflation%\thanks{Grants or other notes
}
%\subtitle{A B-Polarization Satellite Mission \\
%for Detecting Primordial Gravitational Waves \\
%Generated During Inflation}

\author{Paolo de Bernardis, Martin Bucher, Carlo Burigana and Lucio
Piccirillo\\ (for the B-Pol Collaboration)${}^{*}$}

\institute{${}^{*}$ See http://www.b-pol.org for the full list of collaboration members
and a full copy of the B-Pol proposal. See also section 5.
%; Corresponding author:
%Paolo de Bernardis, Dipartimento di Fisica, Universit\'a La Sapienza,
%P.le A. Moro 2, 00185 Roma, Italy, ph. +39 0649914271
}

\date{Received: January 7, 2008 / Accepted: August 4, 2008 - Experimental Astronomy \\ The original publication is available at www.springerlink.com }

\maketitle

\begin{abstract}

B-Pol is a medium-class space mission aimed at detecting the
primordial gravitational waves generated during inflation
through high
accuracy measurements of the Cosmic Microwave
Background (CMB)
polarization. We discuss the scientific
background, feasibility of the experiment, and
implementation developed in response to the
ESA Cosmic Vision 2015-2025 Call for Proposals.
\end{abstract}

\def\lsim{\,\lower2truept\hbox{${< \atop\hbox{\raise4truept\hbox{$\sim$}}}$}\,}
\def\gsim{\,\lower2truept\hbox{${> \atop\hbox{\raise4truept\hbox{$\sim$}}}$}\,}

\section{B-Pol Science}

The quest to understand the origin of the tiny fluctuations about a
perfectly homogeneous and isotropic universe lies at the heart of both
modern cosmology and high-energy physics. Inflationary theory offers
the most satisfying and plausible explanation for the initial
conditions of the universe. Inflation is a phase of superluminal
expansion of space itself, within $10^{-35}s$ of the Big Bang, during which
quantum fluctuations are stretched to cosmological scales
%(see e.g. \cite{mukh81}, \cite{guth82}, \cite{ruba82},  \cite{star82}, \cite{lind83}, \cite{abbo84},
% \cite{kolb90}).
(see e.g., %\cite{mukh81,guth82,ruba82,star82,lind83,abbo84,kolb90}).
\cite{star82,mukh81,guth82,hawk82,ruba82,lind83,bst84,abbo84,kolb90}).
Results from Cosmic Microwave Background (CMB) experiments
have established that the universe is almost
spatially flat, with a nearly Gaussian, scale-invariant spectrum of
primordial adiabatic perturbations
(see e.g., \cite{debe00,stom01,sper03,sper06}).
These features are all consistent with the simplest models of inflation.
By providing accurate measurements of the E-mode (gradient component)
polarization of the CMB, the ESA mission Planck will offer more
stringent tests of the inflationary paradigm \cite{Plan05}.
Nevertheless, even with
such an accurate characterization of the scalar perturbations, a
decisive confirmation of inflation will be lacking and large
uncertainties in the allowed inflationary potentials will persist.
Inflation predicts the existence of primordial gravitational waves on
cosmological scales.
Their detection would firmly establish the
existence of a period of inflationary expansion in the early universe,
and confirm the quantum origin of cosmological fluctuations that led to
the large scale structure observed today. The search for primordial
B-mode (curl component) polarization of the CMB provides the only
opportunity to detect in the foreseeable future the imprint of these
gravitational waves. Measuring the amplitude of these tensor perturbations at one length
scale would fix the energy scale of inflation and its potential.
Measuring their amplitude at more than one length scale would provide a
powerful consistency check for a broad class of inflationary models.

\begin{figure}[ht]
\begin{center}
\vskip 2.cm
{
\setlength{\unitlength}{0.6cm}
\begin{picture}(14, 6)(0,0)
\put(11.6,7.2){TT,scalar}
\put(11.6,6.3){TE,scalar}
\put(11.6,5.8){EE,scalar}
\put(11.6,4.3){BB$\leftarrow $ EE, scalar lensed}
\put(11.6,3.9){{\tiny TT, tensor $(T/S)=10^{-1}$ }}
\put(11.6,3.0){{\tiny TE, tensor $(T/S)=10^{-1}$ }}
\put(11.6,2.75){{\tiny EE, tensor $(T/S)=10^{-1}$ }}
\put(11.6,2.4){BB, $(T/S)=10^{-1}$}
\put(11.6,1.7){BB, $(T/S)=10^{-2}$}
\put(11.6,1){BB, $(T/S)=10^{-3}$}
\put(5.,0.2){Multipole number ($\ell $)}
\put(0.6,3){\begin{rotate}{90} $\ell (\ell +1)C_{\ell }/2\pi ~[\mu K^2]$ \end{rotate}}
\includegraphics[width=7.2cm]{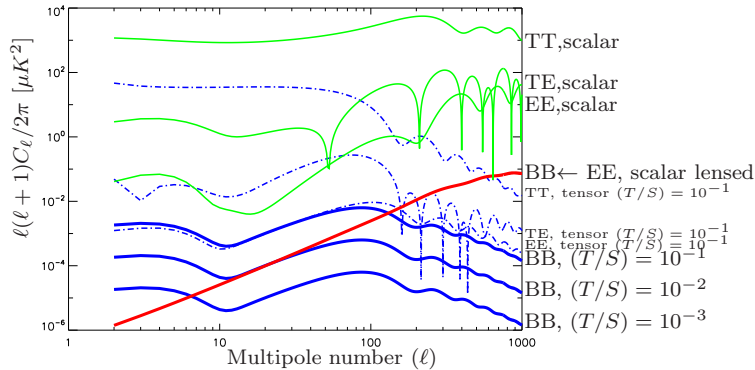}
\end{picture}
}
\end{center}
\vskip -0.5cm
\caption{\small \baselineskip=8pt {{\bf Inflationary Prediction
for the CMB Temperature and Polarization Anisotropies, for the
Scalar and Tensor Modes.} The horizontal axis indicates the
multipole number $\ell$ and the vertical axis indicates $\ell
(\ell +1)C_\ell ^{AB}/(2\pi )$ in units of $(\mu  K)^2$, which is
roughly equivalent to the power spectrum per unit of $\ln\ell$.
The green curves indicate the TT, TE, and EE
power spectra (from top to bottom) generated by the {\it scalar}
mode assuming the parameters from the best-fit model from WMAP
three-year data.
The BB scalar component (indicated by the heavy red
curve) results from the gravitational lensing of the EE polarized
CMB anisotropy
by structures situated mainly around redshift $z\approx 2$. The top
four blue curves (from top to bottom on the left) indicate the TT,
TE, BB, and EE spectra (BB is the heavy solid curve) resulting from the
{\it tensor} mode, assuming a scale-invariant ($n_T=0$) primordial
spectrum and a tensor-to-scalar ratio $(T/S)$ of $0.1$. This value
corresponds roughly to the upper limit established by WMAP. The
bottom two blue curves indicate the tensor BB spectrum for $(T/S)$
equal to $0.01$ and $0.001$, respectively. For the TE cross-correlations
we have plotted the log of the absolute value.} }
\label{Fig:UnderlyingAnisotropies} %1
\end{figure}

\begin{figure}[th]
\includegraphics[width=9.cm,height=10.5cm,angle=90]{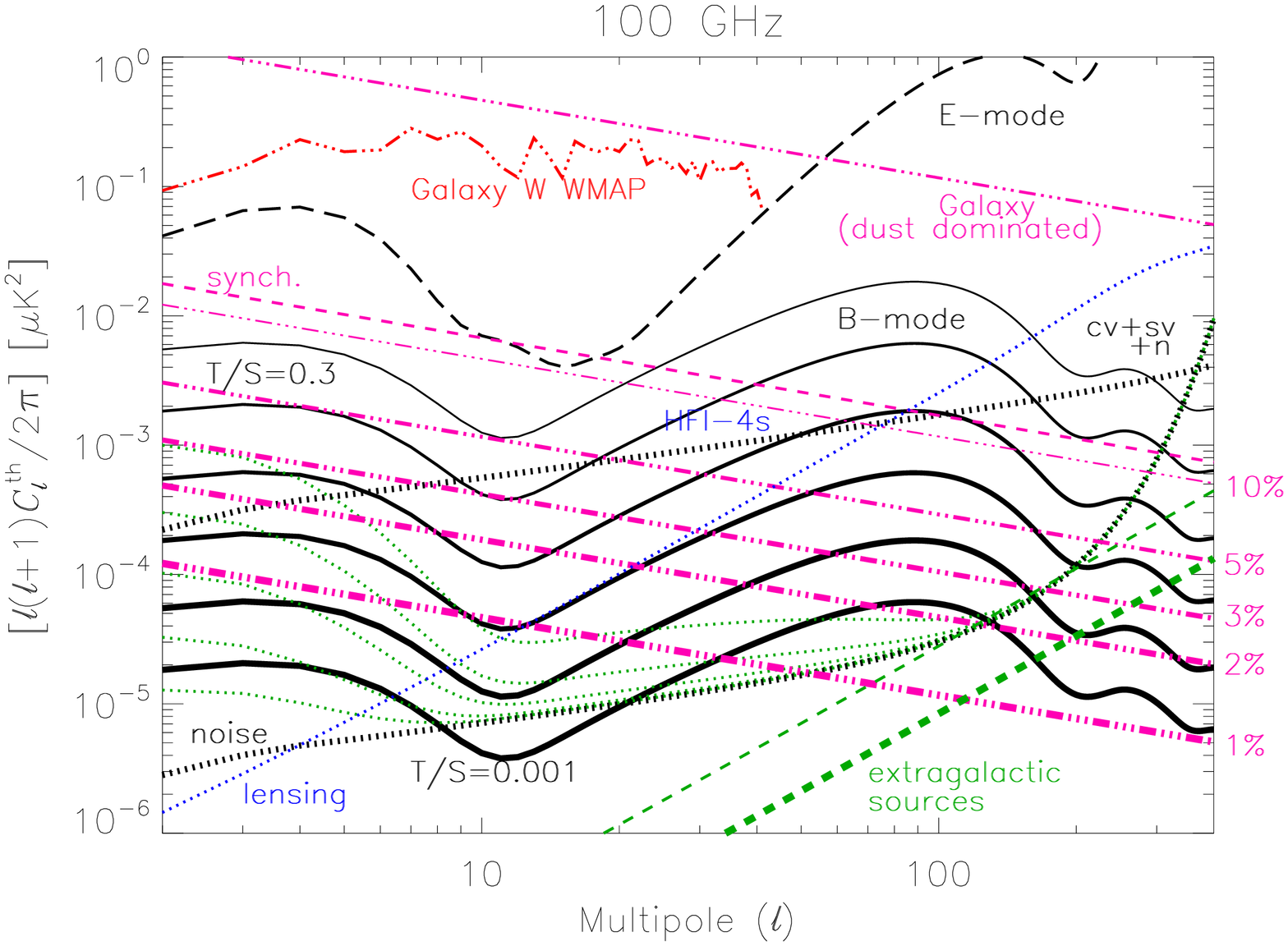}
\caption{\small \baselineskip=8pt
{\bf B Mode Signal and Foregrounds.}
CMB E and B polarization modes compatible with WMAP 3-yr data are compared to Galactic
and extragalactic polarized (B-mode) foregrounds. The anticipated foreground
subtraction
residuals are compared to the B-mode induced by lensing (blue dots)
and to the B-Pol sensitivity. The plots include
cosmic and sampling variance plus instrumental
noise (green dots labeled with
cv+sv+n; black thick dots, noise only)
assuming a multipole binning of 30\%.
The B-Pol frequency channel at 100~GHz is considered here.
The corresponding Planck-HFI instrumental noise
sensitivity is also displayed for comparison (four surveys, upper black thick dots).
The E mode is denoted by the black long dashes. The B mode (black solid lines)
is shown for $T/S = 0.1, 0.03, 0.01, 0.003, 0.001$ from top to
bottom, at increasing thickness. Note that the
cosmic and sampling (74\% sky coverage) variance implies a
dependence of the overall sensitivity at
low multipoles on $T/S$ (again the green dots
refer to $T/S =$ 0.1, 0.03, 0.01, 0.003, 0.001 from top to
bottom),
which is relevant for parameter estimation; instrumental noise only
determines B-Pol's capability to detect the B mode.
Galactic synchrotron (purple dashes) and dust (purple dot-dashes) polarized
emissions produce the overall Galactic foreground (purple three dot-dashes)
that is dominated by dust at 100 GHz. WMAP 3-yr power-law
fits for uncorrelated dust and synchrotron have been
used. For comparison, WMAP 3-yr results derived directly from the foreground
maps are shown on a suitable multipole range: power-law fits provide (generous)
upper limits for the power at low multipoles.
Residual contaminations by Galactic foregrounds
(purple three dot-dashes) are shown for 10\% to 1\% of the map level,
at increasing thickness, as labeled on the right.
The residual contribution by unsubtracted extragalactic sources,
$C_\ell^{\rm resPS}$, and the corresponding uncertainty, $\delta C_\ell^{\rm resPS}$,
computed assuming a relative uncertainty
$\delta \Pi / \Pi = \delta S_{\rm lim}/S_{\rm lim} = 10$\%
in the knowledge of their degree of polarization and in the determination of the source
detection threshold, are also plotted as green dashes, thin and thick, respectively.
}
\label{fig:science}
\end{figure}

If as suggested by recent CMB and large scale galaxy surveys, the power
spectrum of primordial perturbations is not exactly scale invariant,
then in a wide class of inflationary models the level of gravitational
waves will be within the range accessible to a properly designed
mission, as shown in Fig.~\ref{Fig:UnderlyingAnisotropies}.
The bulk of the statistical weight for
detecting inflationary B modes is concentrated at two angular scales on
the sky (see Fig.~\ref{Fig:UnderlyingAnisotropies}):
firstly, at the reionization bump at multipoles $\ell = 2-10$ and
secondly at the multipole region from about 20 to 100 (corresponding to
angles larger than $\approx 1^\circ $ on the sky).
Given that most of the
signal lies on large angular scales, a full-sky survey with exquisite
stability and control of systematic errors of both instrumental and
astrophysical origin is required, hence the need to go to space.
The B mode polarization is a clean probe of gravitational waves, since primordial
scalar perturbations do not contribute to B-modes, and the effects due
to intervening gravitational lenses are calculable and of order
$5 \mu K \cdot {\rm arcmin}$.
This sets the sensitivity target for the mission. In fact, as shown by the blue dotted
line in Fig.~\ref{fig:science}, the lensing contribution to the B-mode
is below the primordial B-mode for tensor to scalar perturbation ratios, $r=T/S$, above
few~$\times 10^{-2}$ at least at multipoles $\ell \lsim 100$, while, for example, a lensing
subtraction at $\sim 10$\% accuracy level (in terms of angular power spectrum) is adequate to
identify the primordial B-mode for $T/S$ above few~$\times 10^{-3}$.

\def\gtorder{\mathrel{\raise.3ex\hbox{$>$}\mkern-14mu
             \lower0.6ex\hbox{$\sim$}}}
\def\ltorder{\mathrel{\raise.3ex\hbox{$<$}\mkern-14mu
             \lower0.6ex\hbox{$\sim$}}}
\newcommand{\muK}{\mu  {\rm K}} \newcommand{\muKarcmin}{\mu  {\rm K\cdot  arcmin}}
\newcommand{\kmbysbyMpc}{\,{\rm km~s^{-1}~Mpc^{-1}}}

A confirmation of inflation and determination of the inflationary potential would have
profound implications for fundamental physics by providing new
experimental data on the physics near the Planck scale. The constraints established would be
indispensable for model building in string and M theory.
The energy scales probed by polarization measurements
lie many orders of magnitude beyond any
conceivable accelerator experiment. Consequently, the quest for
primordial gravitational waves from inflation constitutes a unique
window for constraining the new physics near the Planck scale, which
will help understand how quantum gravity unifies with the other
fundamental interactions.

The implementation of this mission requires significant advances
in three main areas: (1) a sensitivity to tensor modes
of a factor of about 100 with respect to Planck, (2)
control of systematic effects at the level of a few nK,
and (3) a precise ($\sim 1$\%) knowledge of the galactic
foreground polarization.

Aspects (1) and (2), related to the B-Pol design,
will be extensively discussed in the following sections.
Concerning (3), Fig.~\ref{fig:science} compares the
CMB B-mode (for various $T/S$ values) to the B-mode expected from the most relevant
polarized foregrounds and their potential residuals assuming different
levels of accuracy for their subtraction assuming B-Pol sensitivities.
As is evident, a removal of the foreground, and in particular of the Galactic emission,
at the level of about 1\% is necessary to
detect the primordial CMB B mode for $T/S \sim 10^{-3}$.
Foreground subtraction to this level can already be
achieved exploiting already (or soon)
available all-sky (or large area sky coverage) surveys.
Important new data sets will become available
%Substantial improvements will take place are expected
in the next years regarding
polarized foregrounds in the microwave
(e.g., further WMAP surveys, QUAD, BICEPT, EBEX, BOOMERanG, QUIET, C$\ell$OVER, SPIDER, Planck, etc.),
radio (e.g. S-PASS, PGMS, C-BASS, GEM)
and far-IR (e.g. PILOT) bands.
In parallel, it will be crucial to generalize to
polarization the component separation methods successfully applied to
temperature data
(e.g. Wiener filtering, maximum-entropy, Spectral-Matching ICA,
CCA, phase methods) as well as to refine the already existing component
separation methods for polarized data (e.g. template fitting methods, ICA, FastICA, PolEMICA),
and to develop new techniques.

Obviously, the broad frequency coverage proposed for B-Pol
is crucial to allow the application of these methods
at the required level of sensitivity.

Finally we remark that because of its high sensitivity and
accuracy in polarization, a mission devoted to B-modes would make substantial
contributions in several other areas of astrophysics,
such as the physical modeling
of Galactic magnetic fields,
interstellar dust and gas properties including turbulence effects
\cite{cho_lazarian}, and of cosmology, such as gravitational lensing
of the CMB, cosmological reionization, and magnetic fields in the early
universe
(see e.g. respectively \cite{lewis_challinor,lesgourgues_etal}, \cite{burigana_etal},
\cite{scannapieco_ferreira}, and
references therein).

\section{The B-Pol Instrument}
\label{instrument}
B-Pol is a medium class satellite with broad frequency coverage to
enable reliable removal of Galactic foreground contamination and an
angular resolution good enough to access both multipole windows for
detecting primordial gravitational waves from inflation. Accessing the
first window requires a
full-sky survey, possible only from space. The required sensitivity
would nominally require more than 100 years of
integration time for the ESA-Planck mission, and moreover
much better control of systematics
in polarization.

To reach the required instrument performances, the detector
sensitivity $s_{det}$ of 50 $\mu K\surd s$ can be achieved for an
overall instrumental efficiency of 0.5 and a total bolometer NEP
(Noise Equivalent Power) of typically $8\cdot10^{-18} W/\surd Hz$,
close to background limited performance. The sensitivity goal in the 6
bands requires a large number of pixels and a long duration
mission. Typical values are a total of 2000 detectors for a
mission duration of 2 years. The resulting baseline instrumental
configuration is summarized in Table \ref{bpol_char}.
\begin{table}
\begin{center}
\begin{tabular}{lrrrrrr}
  \hline
  Freq. band (GHz)                   & 45     & 70    &   100 &   143 &    217 &    353 \\
  \hline
  $\Delta \nu$                       & 30\%   & 30\%  & 30\%  & 30\%  & 30\%   & 30\%   \\
  ang. res.                          & 15deg  & 68'   & 47'   & 47'   & 40'    & 59'    \\
  \# horns                           & 2      & 7     & 108   & 127   & 398    & 364    \\
  det. noise ($\mu K \cdot \surd s$) & 57     & 33    & 53    & 53    & 61     & 119    \\
  Q \& U sens. ($\mu K \cdot$ arcmin)& 33     & 23    & 8     & 7     & 5      & 10     \\
  Tel. diam. (mm)                    & 45     & 265   & 265   & 185   & 143    & 60
    \\
  \hline
\end{tabular}
\end{center}
\caption{Main characteristics of the B-Pol instrument}
\label{bpol_char}
\end{table}
For a target sensitivity of $r=10^{-3}$, the r.m.s. signal in
primordial B-modes is around 10~nK, and rejecting parasitic
signals to better than this level imposes very stringent requirements
on the polarimeter design and calibration. The strategy for
satisfying these requirements is to combine a very stable environment with a
carefully designed scan and modulation scheme including
redundancies at multiple timescales. In particular, by modulating
the polarization signal with a half-wave plate, both $Q$ and $U$
from a given sky pixel can be measured by a single instrument
pixel on timescales short compared to the detector 1/$f_{\rm
knee}$. If we demand that the residual systematics in Stokes maps
after correction are less than 10\% of the expected signal from
primordial B-modes for $r=10^{-3}$, the uncorrected instrumental
polarization (i.e. conversion of total to linearly polarized
intensity) due to the polarimeter must be below $10^{-5}$, and the
cross-polarization ($Q$ and $U$ mixing) below $5\times 10^{-4}$.
The latter corresponds to a mis-calibration of polarization angle
$< 0.03^\circ$.  Assuming marginalization over absolute
calibration errors during parameter fitting, a 5\% uncertainty
increases the random error on $r$ by only 10\% (increasing to 60\%
if only modes with $\ell > 20$ are used). These requirements can
be relaxed by additional instrument rotation from a well-chosen
scan strategy, as explained below.
The experience so far acquired
from sub-orbital experiments suggests that the optimal experiment
would combine the purity of radiometer front-ends and the high sensitivity of bolometers.
The B-Pol receivers follow this concept.

\begin{figure}
\begin{center}
  \begin{tabular}{cc}
    \includegraphics[width=1.3in]{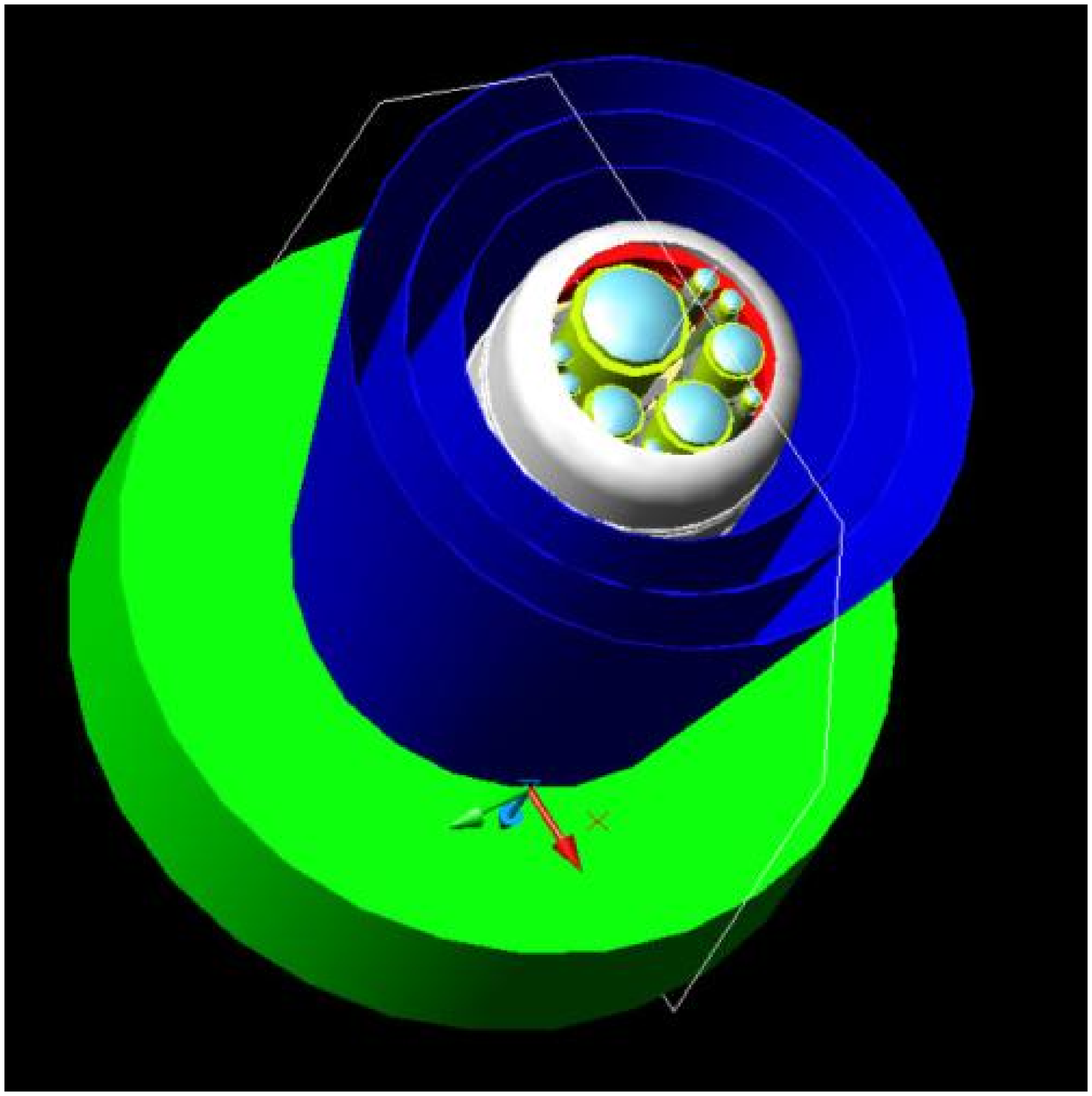}
       \includegraphics[width=1.1in]{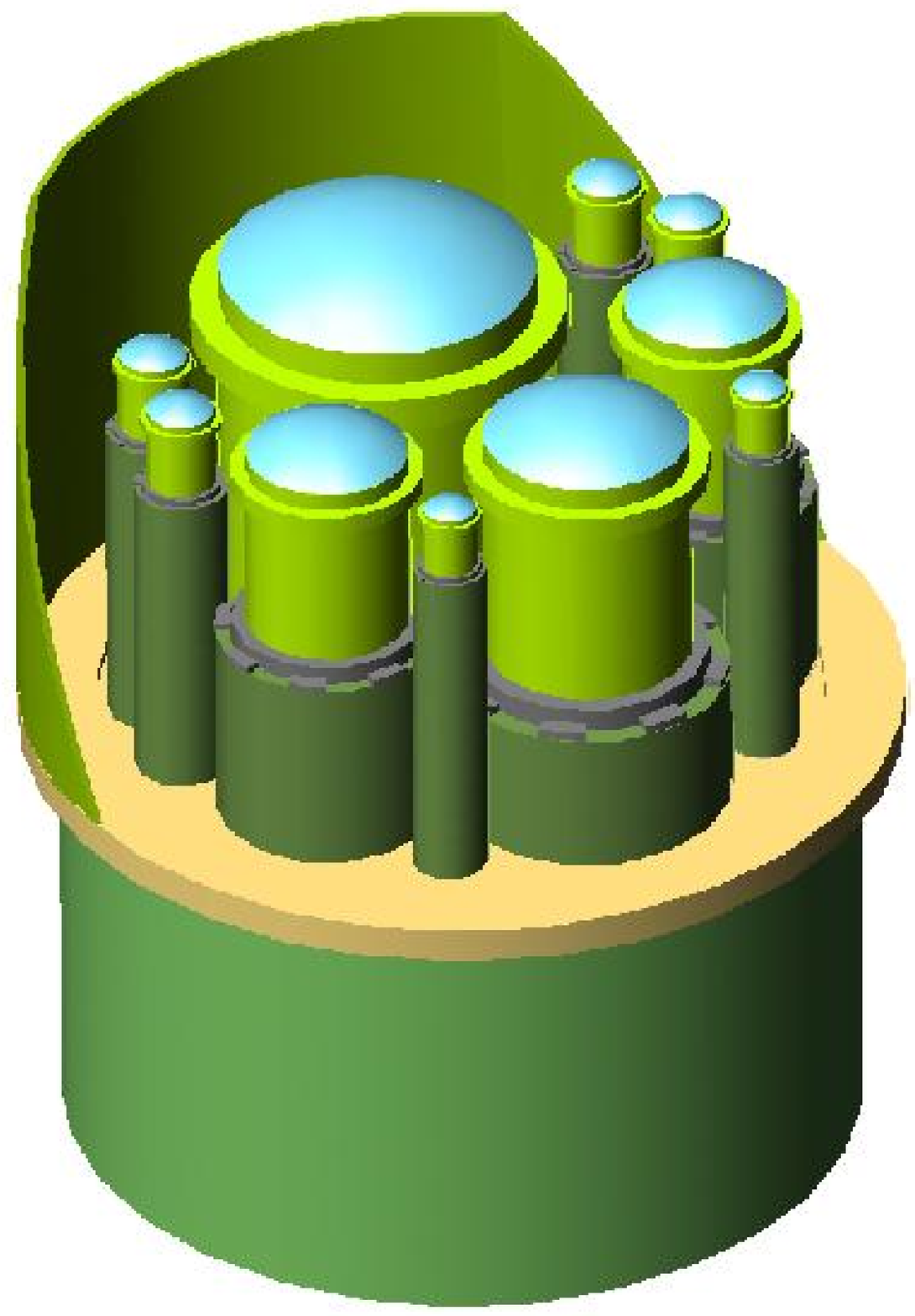}
 \includegraphics[width=2.4 in]{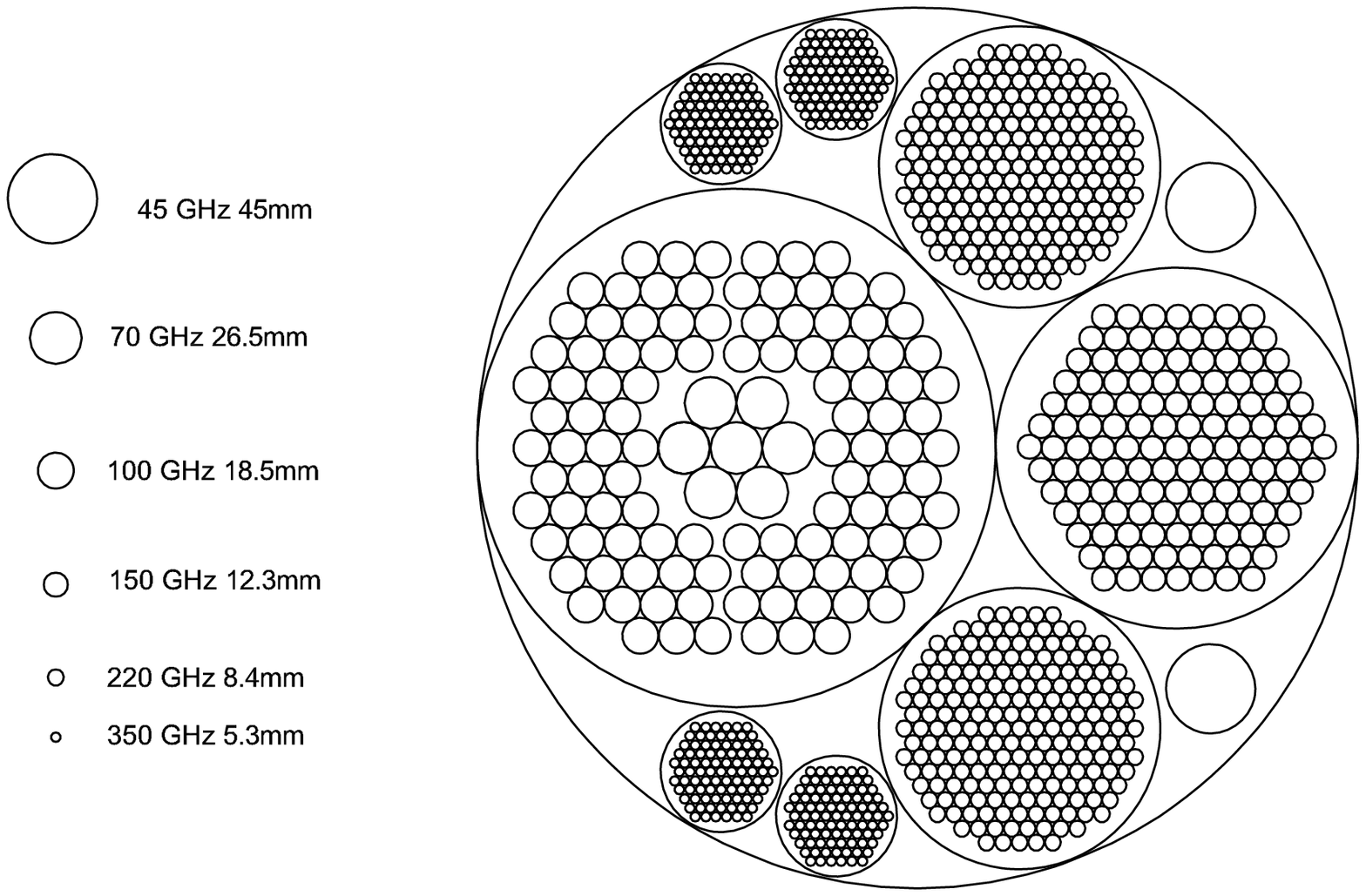}
  \end{tabular}
\end{center}
\caption{An artist's conception of the B-Pol satellite (left), of the
cryogenic instrument (center), and of the details of the focal plane
arrays (right). The cryogenic instrument is enclosed in a 62 cm
diameter, 39 cm height cylinder, shown here on top of the cylinder
enclosing the sub-K refrigerator unit and the cryogenic readout
system, which has similar dimensions. The satellite fits in the
bay of a Soyuz launcher.} \label{bpol}
\end{figure}

A possible implementation of the B-Pol instrument is composed of
8 small telescopes
co-aligned with the spacecraft axis. In each telescope's focal
plane there is an array of single mode corrugated feed-horns
designed to be well matched with the optics and with minimal
aberrations. This configuration is sketched in fig. \ref{bpol}. In Table 2 we report a breakdown of power, volume and mass for the B-Pol instrument.

{\bf Polarimetry:} The incoming polarized radiation is modulated
by a single quasi-optical Half-Wave Plate that rotates in front of
a whole array. Every array pixel consists of a corrugated horn
coupled to an Orthomode Transducer (OMT). The OMT cleanly splits the two incoming
orthogonal polarizations that are then detected by two TES
detectors. OMT and planar phase switches can be fabricated,
allowing for intrinsic on-chip polarimetry and therefore eliminating the
need for a Half-Wave Plate (HWP) provided that this development is successful during
B-Pol phase A. The OMTs are constructed in the waveguide at low frequencies
(45 and 70~GHz) \cite{Boif91} and in a planar microstrip at the higher frequencies
($\ge 100$~GHz). $Q$ and $U$ are extracted by differentiating the
detector outputs at different HWP angles.
The 45~\&~70~GHz receivers feature a variant of this approach based
on a pseudo-correlator scheme, using bolometers as the final
detectors.

An achromatic HWP can be made by stacking together %more
several birefringent
plates oriented in different directions following the Pancharatnam
recipe \cite{pisa06} \cite{savi06}. Cross-polarizations of
order of -30~dB are easily achievable in wide bands using 3
plates.
An alternative achromatic HWP uses the
same photo-lithographic techniques
adapted for submillimeter mesh
filters in order to fabricate artificial birefringent materials.

A special
mechanism is necessary to rotate the HWPs, producing
modulation of polarization. To achieve a smooth, continuous
rotation, each HWP is mounted on a hollow magnetic bearing.
These devices have already been used in space (in supporting
flywheels for attitude control systems).
The rotation is induced by custom made spin-up
superconducting motors similar to the type %one
described in
\cite{hana03}.

A completely symmetric planar OMT can be built using four
probes inside a circular waveguide which are then recombined.
Broadband four-probe antennas OMT have been built and tested in
the L-band. For a 30\% bandwidth the measured return loss and
cross-polarization were, respectively, around 20~dB and -40~dB
\cite{enga03}. For the W and D bands niobium planar antennas are supported
by a very thin SiN layer that is held by a thicker Si substrate
etched away
to fabricate the waveguide and the probes.
The return loss and the cross polarization across the bands are
less than 25~dB and -50~dB, respectively.
Broadband waveguide OMTs are used for the two lower frequencies 45 and
70~GHz. Return loss and isolation of order 20~dB and 50~dB,
respectively, have already been achieved at 100~GHz for a 30\%
bandwidth \cite{pisa07}, and 40 and 70~dB at 30~GHz \cite{peve06}.

{\bf Detectors : } The baseline requirement
is to achieve at all wavelengths
high-fidelity polarimetry over 30\% bandwidths with
NEPs of $8\cdot10^{-18} W/\surd Hz$ in each polarization, which
ensures a CMB photon noise limited sensitivity. Superconducting
Transition Edge Sensors (TESs) are ideal for this purpose. TESs
have been developed extensively for astronomical observations at
millimeter through x-ray wavelengths and have a long history
of successful use.
Systems using various materials can be combined, and through the proximity effect,
bilayers (e.g., MoCu, MoAu) chosen to match the power handling and cooling
requirements of the instrument.
For B-pol we have selected microstrip-coupled TESs, where a
thin-film waveguide probe is used to couple radiation from a horn
antenna onto the TES through a superconducting microstrip
transmission line terminated with a resistor. With this
configuration, corrugated horns can be used to achieve high optical
efficiency with low cross polarization, with the significant
advantage that it is
possible to calculate precisely the optical performance of the
whole instrument. Superconducting planar bandpass filters can be
lithographed onto the detector chips, and planar OMTs can be
fabricated allowing for intrinsic on-chip polarimetry.
Microstrip-coupled TESs have already been demonstrated in the laboratory by a
number of groups, and we have chosen as our baseline a
four-probe, membrane-based architecture of the kind being
developed for C$\ell$OVER \cite{audl06}. A very recent measurement on a
single pixel of C$\ell$OVER has achieved an optical efficiency higher
than 90\%.

Any instrument designed for CMB polarization studies requires a high
($> 80$\%) in-band transmission of the filter stack and simultaneous
rejection of optical/NIR power better than
$\sim 10^{-12}$. Re-using know-how developed for the Planck HFI
makes it
possible to develop a filter stack satisfying these stringent
requirements.

{\bf Optics : } Specifications directly relevant to the optical
design are $\sim$ 50 arcmin resolution, 1\% spillover, $\sim$ 1\%
beam ellipticity, and a 30dB maximum cross-polarization. While both
reflective/mirror and refractive/lens telescopes could be used
to achieve the resolution, a lens-based design has been chosen to
minimize the required cold volume for the instrument. To overcome
the chromaticity problems that lenses and half-wave plates could
cause with our wide frequency span and in order to achieve low
aberration, we have sub-divided the payload into 8
telescope/focal plane systems. Each telescope covers a maximum of 2
frequency bands of 30\% bandwidth each. A 3-lens, F/1.8
configuration has been demonstrated by optical ray-tracing
to achieve the required performance in terms of
cross-polarization, beam symmetry, and size of the usable focal
plane. In order to reduce the risk of stray light and
systematic effects, each focal plane is populated by $\sim
6\lambda$ corrugated feed horns, their phase center located
at the focal surface of the telescope. Corrugated horns are
extensively used in CMB experiments due to their low sidelobes and
very high polarization purity (see e.g.
\cite{vill02},\cite{maff04}). The diameter of the feeds is chosen
to minimize edge diffraction and spillover (i.e. -30dB of edge
taper for F/1 optics). Modules of 7 horns in a single hexagonal
module are used to reduce cost and to build the entire
focal plane in a simple way. Different manufacturing methods are available:
electroforming, direct machining and platelet structure
\cite{kang05}, \cite{haas94}. The horns form a curved focal
surface to optimize the illumination of the lens system and reduce
aberration. Moreover, each optical system is located in a 4K
black enclosure (eccosorb-like) to make the optical power coming
from the 1\% spillover totally un-polarized. This optical concept
is valid for all spectral bands but one. Due to the
limited size of the payload, we propose to have only 2 channels at
45GHz with a limited resolution of 15 deg for foreground
separation. These two low frequency channels serve as
a full-sky calibration reference to merge with smaller but
deeper ground-based surveys, which is carried out efficiently
below 60 GHz. Two stand-alone corrugated conical horns without resort to
lenses is optimized to reduce sidelobes and cross-polar
components.

{\bf Cryogenics : }

The B-Pol cryogenic system uses the heritage from ISO, Planck, and
Herschel European space missions, more specifically, the V-groove
passive cooling system, lHe cryostat, and sub-Kelvin coolers.
This setup should provide a cooling power of 1 $\mu$W at 100-150 mK for
the detectors and a cooling power of 800 $\mu$W at 2 K for
the cold electronics and the optics as well as active or passive
regulation of the various thermal stages so that the induced
systematic effects contribute negligibly to the scientific signal error
budget. The cryogenic chain must
be compliant with pre-launch operations, withstand
launch vibrations, and have overall mass/volume/power consumption
compatible with launcher and service module.

In addition to an outer screen that provides a first shielding from
the Sun, a {\it passive radiator} provides a low-cost
initial cooling stage for a wet cryostat.
For B-Pol, the power must be radiated along the
viewing line-of-sight, which covers almost half the celestial
sphere within a few days, i.e., an orientation from 90 to 180 degrees
away from the Sun. An industrial study was carried out for the Sampan
CNES proposal.
With forward V-grooves covering an
effective area of a few $m^2,$ we
can achieve temperatures of less than 50 K at the last internal
V-groove. Unlike the mechanical coolers used in Planck,
a {\it wet cryostat} provides a more compact and lighter solution,
and does not produce vibration and magnetically induced parasitic
effects on the detectors. The heritage from ISO and Herschel is
important in that respect.
Either liquid superfluid
helium (lHe) or solid hydrogen (sH) can be used.
In the Sampan case, an unoptimized configuration
(compatible with V-grooves at 80 K)
with a 2.5 year lifetime consists of
typically 88 kg of lHe for an overall cryostat mass of 352 kg, or
21 kg of sH for an overall cryostat mass of 109 kg. Both
cryostats have similar dimensions. Two possible options for the
{\it sub-K cooler} are already at a readiness level suitable for
this mission. The heritage from Planck HFI 3He-4He open-cycle
dilution system is direct.
The system has no magnetic influence on the detectors. A powerful
passive isolation of the bolometer array plate can be achieved, as in
HFI, by using HoY alloys. It can also provide a Joule-Thomson
intermediate stage at 1.6 K, which is useful in the case of a sH
cryostat.
Adiabatic demagnetization of a paramagnetic salt has
some possible advantages over the dilution technique, dispensing with
the
need for liquid confinement and providing a larger cooling power.
The main disadvantage is the need of a strong magnetic field
obtained with a superconducting magnet, but such a field can be fully
contained by means of a superconducting shield. The problems
arising for space applications have already been solved.
Adiabatic demagnetization refrigerators with
an operating temperature below 0.1K have been deployed on rockets
for X-ray microcalorimetry.

\begin{table}
\begin{center}
\begin{tabular}{lrrrr}
\hline Subsystem & mass (Kg) &  volume (liters) & power (W) \\
\hline {\bf total sub-K cooler}        &  {\bf 20} & {\bf 80} & 10
\\ {\bf total focal plane}         &  {\bf 74} & {\bf 80} &  - \\
horns and focal plane     &   29      &          &    \\
telescopes & 29 &          &    \\ rotation mechanisms       & 16
& &    \\ {\bf total cryo harness}        &  {\bf 5}  & {\bf 20} &
\\ DBU &  13.2 & &
\\ BEU                       &   9.6     &          &    \\ DHU
&   6.6     &          &    \\ MMU                       &   9.6 &
&    \\ PSU                       &   6.6     & &    \\ WPU & 10.1
&          &    \\ warm harness              & 9.6 & &    \\ {\bf
total warm electronics}    &  {\bf 65} & {\bf 40} & 190\\ \hline
{\bf grand-total instrument}    &  {\bf 164}& {\bf 215}&{\bf 200}
\\ \hline
\end{tabular}
\end{center}
\caption{Breakdown of B-Pol instrument resources: mass, volume,
power. Masses and powers include 20\% contingency. The science
data produced by the instrument amount to 0.64 Mbit/s, including a
mild compression factor. } \label{tab:reso}
\end{table}

\section{The B-Pol mission}
\label{mission}

The rejection of parasitic signals from the side lobes, together
with the need of a full sky coverage and the need to avoid Sun,
Earth and Moon emission, led to choosing an L2 orbit for
Planck. The same considerations apply to B-Pol.
A Soyuz launcher that would
take off from Kourou offers the most cost effective solution compatible with
the instrument requirements. A Lissajous orbit around L2 is preferred
to a halo orbit because it is easier to control. It can be phased
to avoid eclipses for the entire duration of the mission and
can cope with larger payload masses. The mission would last 2 to 4
years depending on the actual cryogenic performance. The transfer
from Earth to L2 would last 2 to 4 months. The Calibration and
Performance Verification Phase would last 2 months. To achieve
optimal sky coverage and redundancy (both in terms of hits per
pixel and angular coverage), B-pol uses a complex scan that
consists of a precession about the anti-solar axis and a nutation
about this precession axis, very much like the WMAP mission. These
two motions, with adequately related
rotation periods ensure a
large sky coverage ($\sim 50\%$) over a short timescale ($\sim
2$~days) which together with a shorter ($\sim$~seconds)
polarization modulation is optimal for $1/f$ noise and
instrumental drifts rejection. Inflight calibrations are required
to correct for gain, polarization angle, differential gain of
bolometer pairs, and contamination factors for each receiver ($T$
into $Q$~\&~$U$; $Q$ and $U$ mixing).
Such calibration is achieved using a
set of polarized and unpolarized sky
sources observed throughout the entire mission.
Onboard artificial sources are also envisaged for more frequent
calibration transfer.

\section{Conclusions}
\label{conclusions}
The search and discovery of primordial B modes provide a unique
window for exploring new physics near the Planck scale.
There are no competing experimental probes able to access this
domain. We have presented an implementation of a CMB polarization
mission called B-Pol that fits within the tight requirements
for medium-size missions within the ESA Cosmic-Vision 2015-2025
program. The instrument would survey the sky from the Lagrangian
point L2 of the Sun-Earth system for two years, and produce
maps of the {\it polarized} microwave background anisotropy
with sensitivity
to $(T/S)$ two orders of magnitude better than Planck. B-Pol would explore the
entire parameter space spanned by the ``large-field''
inflationary models, and consequently is capable to measure
the energy scale of inflation.
If, as suggested by current CMB and large scale
galaxy surveys, the power spectrum of primordial perturbations is
not exactly scale invariant, then in a wide class of inflationary
models the level of gravitational waves will lie well within the
range probed by B-Pol.

\vfill \eject

\section{Main Contributors to the B-Pol Proposal}

\subsection{National coordinators:}

P. de Bernardis (IT), F. Bouchet (FR), E. Kreysa (GE), L. Piccirillo (UK), R. Rebolo (SP)

\subsection{Instrument working group:}

P. de Bernardis, L. Piccirillo, P. Ade, M. Bersanelli, M. De Petris, F.X. Desert, E. Kreysa, L. Kuzmin, B. Maffei, N. Mandolesi, S. Masi, P. Mauskopf, T. Peacocke, F. Piacentini, M. Piat, G. Pisano, R. Rebolo, G. Savini, R. Tascone, F. Villa, S. Withington, G. Yassin

\subsection{Science working group:}

M. Bucher, J. Bartlett, F. Bouchet, C. Caprini, A. Challinor, R. Battye, G. Efstathiou, F. Finelli, K. Ganga, J. Garcia-Bellido, F. Hansen, K. Land, A. Jaffe, S. Matarrese, A. Melchiorri, P. Natoli, L. Popa, R. Stompor, B. van Tent, L. Verde

\subsection{Foregrounds working group:}

C. Burigana, T. Arshakian, C. Baccigalupi, A. Banday, D. Barbosa, R. Beck, J.P. Bernard, G. Bernardi, A. Bonaldi, E. Carretti, D. Clements, P. Coles, G. de Zotti, R. Fabbri, P. Fosalba Vela, J. Gonzales-Nuevo, M. Hobson, L. La Porta, P. Leahy, J.F. Macias-Perez, E. Martinez-Gonzalez, M. Massardi, P. Naselsky, N. Ponthieu, W. Reich, S. Ricciardi, F. Stivoli, L. Toffolatti, M. Tucci, P. Vielva

\end{document}